\documentclass[article]{aa}
\usepackage{txfonts}
\usepackage{natbib}
\usepackage{deluxetable}
\bibpunct{(}{)}{;}{a}{}{,}
\usepackage{graphicx}

\begin{document}

\title{The Opacity of Spiral Galaxy Disks VI:\\
Extinction, stellar light and color.
\thanks{Research support by NASA 
through grant number HST-AR-08360 from the Space Telescope Science 
Institute (STScI), the STScI Discretionary Fund (grant numbers 82206 and 
82304 to R. J. Allen) and the Kapteyn Institute of Groningen University.}
}
\author{B. W. Holwerda \inst{1,2} \and R. A. Gonz\'alez \inst{3} \and P. C. van der Kruit \inst{1} \and Ronald J. Allen \inst{2} }

\offprints{B.W. Holwerda, \email{Holwerda@stsci.edu}}

\institute{Kapteyn Astronomical Institute, 
University of Groningen, 
PO Box 800, 
9700 AV Groningen, 
the Netherlands.
\and
Space Telescope Science Institute, 
Baltimore, 
MD 21218
USA
\and
Centro de Radiastronom\'{\i}a y Astrof\'{\i}sica, 
Universidad Nacional Aut\'{o}noma de M\'{e}xico, 
58190 Morelia, Michoac\'{a}n, 
Mexico}

\date{Received / Accepted}

\titlerunning{Extinction, stellar light and color.}
\authorrunning{Holwerda et al.}

\abstract{
In this paper we explore the relation between dust extinction and 
stellar light distribution in disks of spiral galaxies. Extinction 
influences our dynamical and photometric perception of disks, 
since it can distort our measurement of the contribution of the 
stellar component.
To characterize the total extinction by a foreground disk, 
\cite{Gonzalez98} proposed the ``Synthetic Field Method'' (SFM), 
which uses the calibrated number of distant galaxies seen through 
the foreground disk as a direct indication of extinction. The method 
is described in \cite{Gonzalez98} and \cite{Holwerda05a}. To obtain 
good statistics, the method was applied to a set of HST/WFPC2 
fields \citep{Holwerda05b} and radial extinction profiles were 
derived, based on these counts. In the present paper, we explore 
the relation of opacity with surface brightness or color from 2MASS 
images, as well as the relation between the scalelengths for extinction 
and light in the I band. 
We find that there is indeed a relation between the opacity ($A_I$) 
and the surface brightness, particularly at the higher surface brightnesses. 
No strong relation between near infrared (H-J, H-K) color 
and opacity is found. The scalelengths of the extinction are uncertain 
for individual galaxies but seem to indicate that the dust distribution 
is much more extended than the stellar light.
The results from the distant galaxy counts are also compared to the 
reddening derived from the Cepheids light-curves \citep{Keyproject}. 
The extinction values are consistent, provided the selection effect 
against Cepheids with higher values of $A_I$ is taken into account.
The implications from these relations for disk photometry, M/L conversion 
and galaxy dynamical modeling are briefly discussed.
 }

\maketitle
	
\section{\label{secSBAintro}Introduction}

Dust extinction has influenced our perception of spiral disks since 
the first observations of them. The measurements of disk characteristics, 
such as the central surface brightness ($\mu_0$), the typical exponential 
scale ($r_{typ}$) and the mass-to-light ratio (M/L), are all affected by the 
dust extinction in the photometric band of observation. 
The original assertion by \cite{Holmberg58} that spiral disks are optically 
thin to their stellar light came under scrutiny after the paper by 
\cite{Disney90} and the observational result of \cite{Valentijn90} revealed that they 
were, in fact, practically opaque. The debate quickly culminated in a 
conference \citep{Cardith94}, during which many methods to measure 
the opacity of spiral disks were put forward. Notably, two methods do not 
use the disk's own stellar light for the measurement: the occulting galaxy 
technique \citep{kw92, Andredakis92} and the use of calibrated counts of 
distant objects (the ``Synthetic Field Method'' (SFM), by \cite{Gonzalez98}).
%
Thus far, the following picture of the influence of dust on disk photometry 
has emerged from earlier studies, most of which are based on the inclination 
effect on photometry of a large sample of spiral disks.
\cite{Tully98} and \cite{Masters03} reported that disks are more opaque in the 
blue. Disks are practically transparent in the near infrared \citep{Peletier92, 
Graham01b}, making these bands the best mass-to-luminosity estimator 
\citep{deJong96}. Disks are practically transparent in the outer parts but 
show significant absorption in the inner regions \citep{Valentijn94,Giovanelli94}. 
The radial extent of the dust has been explored using the sub-mm emission 
\citep{Alton98, Davies99, Trewhella00, Radovich01} and edge-on models 
\citep{Xilouris99}. These results indicate that the scalelength of the dust is 
40 \% larger than that of the light. 
The disk's average extinction correlates with the total galaxy luminosity \citep{Giovanelli95,Tully98,Masters03}. And spiral arms are more opaque 
than the disk \citep{Beckman96,kw00a}.This may be attributed to a more 
clumpy medium in the arms in addition to a disk \citep{Gonzalez98}.
\cite{Stevens05} found evidence based on the infrared and sub-mm emission 
from dust for two thermal components of the dust, a warm component associated 
with star formation and a colder component in a more extended disk.

In this paper we explore the relation between the light from a spiral galaxy's 
disk and the opacity measured using the SFM. This paper is organized as 
follows: section \ref{secSBASFM} summarizes the ``Synthetic Field Method'', 
section \ref{secSBASysFX} discusses possible systematic effects in the method 
and sample, section \ref{secSBA} describes the relation between surface brightness 
and average extinction, section \ref{secSBAtype} explores this relation for arm 
and disk regions, and section \ref{secColA} the relation between extinction 
and near-infrared color. In section \ref{secScales}, the scalelengths for light 
and extinction are compared. A brief comparison between Cepheid reddening 
and opacity is made in section \ref{secKeyProj}. The implications for 
measurements involving spiral disks are discussed in section \ref{secML}, 
and we list our conclusions in section \ref{secML}.

\section{\label{secSBASFM}The ``Synthetic Field Method''}

The number of distant galaxies seen through a foreground spiral disk is a 
function of dust extinction as well as crowding and confusion in the foreground 
disk. Distant galaxy numbers were used by several authors to measure 
extinction in the Magellanic Clouds and other galaxies\footnote{See for a 
brief review \cite{Holwerda05a}}.  The ``Synthetic Field Method'' was developed 
by \cite{Gonzalez98} to calibrate an extinction measurement based on the 
number of distant galaxies in a Hubble Space Telescope (HST) image. 
It quantifies the effects of crowding and confusion by the foreground spiral 
disk. The SFM consists of the following steps. First, the number of distant 
galaxies in the science field is identified. Secondly, synthetic fields are 
constructed. These are the original science field with a suitable deep field 
added, which is dimmed to mimic the effects of dust. The added distant 
galaxies in the resulting synthetic field are identified, based on object 
appearance and color. The number of these synthetic galaxies identified 
in the synthetic fields as a function of dimming can then be characterized:

\begin{equation}
\label{eq:SBAN}
A = -2.5 ~ C ~ log \left(N \over N_0 \right).
\end{equation}

\noindent $A$ is the dimming in magnitudes, $N$ the number of synthetic 
galaxies retrieved. $N_0$ is the number of galaxies expected in the science 
field if no extinction were present and $C$ is the parameter of the fit that 
depends on the crowding and confusion of the science field. A series of 
synthetic fields at varying values for $A$ is made to accurately characterize 
equation \ref{eq:SBAN} for every science field in question. From the relation above and 
the number of actual distant galaxies identified in the science field, the 
average extinction in the field can be determined. As the cosmic variance 
causes an additional uncertainty in the original number of distant galaxies 
present behind the foreground disk, the uncertainties in the extinction 
determination are high for individual fields. For a complete discussion of 
the uncertainties of the SFM, see \cite{Holwerda05a}. To combat poor 
statistics, the numbers of distant galaxies in several images are combined, 
based on common characteristics of the foreground disks. \cite{Holwerda05b} 
combined numbers based on radius and Hubble types. In this paper we 
compare the numbers of distant galaxies for image sections of common 
surface brightness and color.

\cite{Gonzalez03} and \cite{Holwerda05d} concluded that the optimal 
distance for the SFM is that of Virgo cluster for the HST instruments. Hence our 
sample of fields is taken for disks at this range of distances.

\section{\label{secSBASysFX}Discussion of systematic effects}

There are two possible sources of systematics in the following results: 
the selection of the sample of foreground galaxies and possible systematics 
in the method itself. The systematics and uncertainties of the method are 
also discussed in detail in \cite{Holwerda05a} but we briefly list possible 
systematics here. The selection of the sample is discussed in \cite{Holwerda05b} 
but the effects of the smaller sample on the new segmentation of distant galaxy 
counts are discussed below. 

\subsection{\label{secSBASysSFM}Systematics of the Synthetic Field Method}

A systematic can creep into the SFM if there is a difference in the object 
identification in the science and the synthetic fields. This was one of our 
main drivers to automate the identification process to the highest possible 
degree. However, a visual check of the candidate objects is still necessary. 
Therefore an observer bias can not be completely excluded. See for an 
in-depth discussion regarding the identification process \cite{Holwerda05a}. 

There are, however, several reasons why we consider the counts calibrated 
sufficiently for any systematics. The radial opacity profile does seem to end at 
zero extinction at higher radii \citep{Holwerda05b}. The radial opacity profile 
agrees with the values from occulting galaxies \citep{Holwerda05b}. And 
there is good agreement by different observers on the counts in NGC 4536 
\citep{Holwerda05a}. The Synthetic Field Method was conceived to calibrate 
any observer bias. If our identification process has resulted in the removal 
of blue distant galaxies together with the HII regions, this has been done 
in the same way for both synthetic and science fields and therefore does not affect 
the derived opacity value. 

\subsection{\label{secSBASysSample}Sample selection effects}

\begin{figure}[!h]
\includegraphics[width=0.5\textwidth]{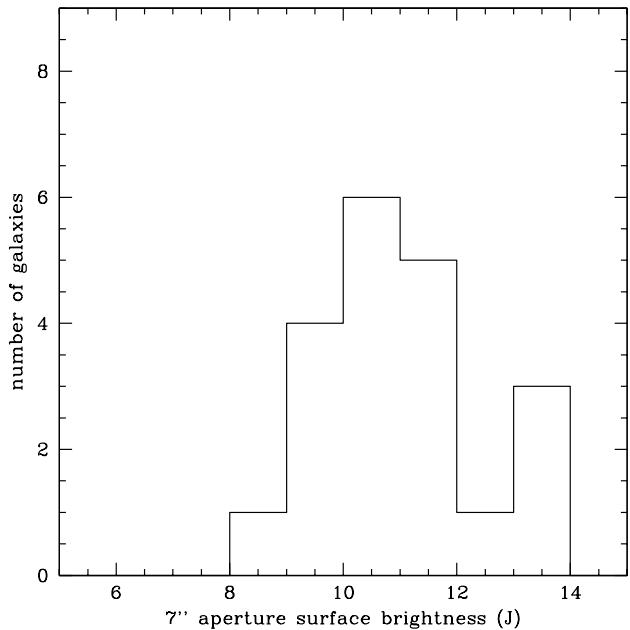}
\caption{\label{MagHist} The number of foreground galaxies as a function 
of 2MASS small aperture (7") magnitude from the Large Galaxy Atlas \citep{LGA}. } 
\end{figure}

Since the SFM requires that we combine several foreground galaxies in 
order to obtain the required statistics on the distant background galaxies, 
there is the risk of selection effects in our results. The selection criteria for the 
data-sets in our sample are described in detail in \cite{Holwerda05b}. 
Essentially face-on spiral galaxies with deep V,I WFPC2 data available 
are selected. To ensure that the galaxy sample does not display two clearly 
separate populations in central surface brightness or color, the histograms 
of central magnitude and morphological type are plotted in Figure 
\ref{MagHist} and \ref{MorphHist}. There is no clear indication for a bimodal 
distribution in the central surface brightness. There is, however, a selection 
effect against the earliest type spirals in our sample as these were not selected 
for the HST Distance Scale Key project.

The opacities from the SFM are not readily corrected for the inclination of 
the disk as this correction depends strongly on the dust cloud morphology 
\citep{Holwerda05b,Holwerda05}. The opacity values can therefore best be 
interpreted as an upper limit of the apparent filling factor of clouds.

From our original sample \citep{Holwerda05b}, the following fields could not be used 
due to problems with the 2MASS fields (M51-2, NGC 4321 NGC 4414-1/2, NGC 4496A, 
NGC 4571, NGC 4603, NGC 4639 and NGC 4725). The two LSB galaxies 
(UGC 2302 and UGC 6614) were excluded as well. The numbers from the 
remaining 23 fields are used in Figures \ref{SB_k_A_0.5} through \ref{Color_J_K_A}. 
There is a spread in morphological types in these galaxies (Figure 
\ref{MorphHist}), a factor to take into account in the interpretation of the following results.


\section{\label{secSBA}Surface brightness and disk opacity}

\cite{Giovanelli94}, \cite{Tully98} and \cite{Masters03} linked the overall 
disk opacity with the total luminosity of a spiral galaxy. It appears that the 
brighter spiral disks are also more opaque. 
The classical relation between gas, dust and stellar mass in the Milky Way 
is often used as a benchmark. If there is a constant ratio between stars and 
dust, some relation is expected between surface brightness and opacity of 
a spiral disk. The relation between 
light and extinction can be explored in more detail, using the SFM. 
\cite{Holwerda05b} compared the average radial opacity of their sample 
to the average radial surface brightness and found a tentative relation 
between the surface brightness of a radial annulus and its opacity based 
on counts of field galaxies (Figure 16 in \cite{Holwerda05b}).

The computation of an average surface brightness per radial interval, 
integrated over all these disks, smoothes all the variations in surface 
brightness. To explore any relation between surface brightness and 
opacity without this smoothing, the counts of distant galaxies must be 
done for a surface brightness interval, not a radial one. 

Each of the distant galaxies found in either synthetic or science fields 
was flagged with the surface brightness of the corresponding position 
in the 2MASS \citep{LGA} image in the H, J and K bands.
This allows us to sort the distant galaxies according to disk surface 
brightness, regardless of their position in the foreground disk. In Figures 
\ref{SB_h_A_0.5}, \ref{SB_j_A_0.5}, and \ref{SB_k_A_0.5} we plot the 
opacity versus surface brightness in 
the H, J, and K band respectively. The top panes show the number of 
field galaxies found at each surface brightness, both in the science 
field and the synthetic fields without any extinction (A=0). 
In Figures  \ref{SB_h_A_0.5}, \ref{SB_j_A_0.5}, and \ref{SB_k_A_0.5} , 
the middle panel shows the opacity without distinguishing for arm and disk. 
The bottom two panels show the derived opacity from the counts in just the 
spiral arm or disk regions respectively.

The drop of the number of synthetic distant galaxies without extinction at higher 
surface brightnesses is a selection effect of the HST data. The majority 
of the WFPC2 images are pointed at the optical disks of the galaxies, 
leaving little solid angle at the lower radii, and hence high surface brightness. 
In addition, crowding effects limit useful solid angle at higher surface brightnesses. 
The limit of the 2MASS photometry is also quickly reached at low surface brightnesses, 
the zeropoints are around 20,5 for both H, K and 21 for J.
As a result, the statistics are sufficient for an opacity measurement 
between approximately 18 to 21 mag arcsec$^{-2}$ in H and K and 
19 and 21 for J.

The surface brightness values were derived from the public 2MASS 
images, as they are relatively uniform and the near infrared emission 
tracks the stellar component of the disk better than other bands.
There are two main concerns in using the 2MASS public images to 
compute surface brightnesses. 
The first concern is whether or not the photometry of them can be compared over a 
series of images and how accurate a surface brightness measurement 
from a single pixel in these images is.
Secondly, significant flux might be contributed by the detected field 
galaxies themselves as a distant galaxy identified in the WFPC2 field 
is about a pixel in size in the 2MASS field. If this is the case, the surface 
brightness measurements of the science field galaxies should be 
generally higher than for galaxies in the synthetic fields, where there is 
no actual distant background galaxy to contribute to the 2MASS flux. 
The first concern can be addressed by simply comparing over large 
surface brightness bins and by using only similar measurements, i.e. the surface 
brightness measurements for the synthetic fields have the same photometric 
uncertainty as the science field ones. The effect of the second concern 
should be evident from the relative distribution of the number of galaxies 
from the science and synthetic fields as a function of surface brightness. 
If there is a systematic offset between these two groups, the distant 
galaxies in the science fields did contribute to the flux. Such an offset of 
the science field objects to the higher surface brightnesses should be 
evident in the histograms in Figures \ref{SB_h_A_0.5}, \ref{SB_j_A_0.5} 
and \ref{SB_k_A_0.5}. None seems to be present. Any such offset would 
work counter to the result of higher opacity with brighter disk surface 
brightness that was found. As an extra check, Figure \ref{SB_k_A_0.1} 
shows the same as Figure \ref{SB_k_A_0.5} but for a smaller bin size in 
surface brightness. In  Figure \ref{SB_k_A_0.1}, an offset is not evident 
as well and the same general trend can be discerned between surface 
brightness and opacity. 

By setting the bin-size in surface brightness much larger than the expected 
surface brightness uncertainty, the scatter in the opacity is significantly 
reduced. For this reason we chose a bin size of 0.5 mag for Figures 
\ref{SB_h_A_0.5}, \ref{SB_j_A_0.5} and \ref{SB_k_A_0.5}.
The uncertainty in the number of distant galaxies depends on the solid 
angle under consideration. The uncertainties in Figures \ref{SB_h_A_0.5}, 
\ref{SB_j_A_0.5}, \ref{SB_k_A_0.5} and \ref{SB_k_A_0.1} are based on 
the total solid angle in the whole of the mosaics with a surface brightness 
value in the interval. A reliable observation could be made for surface 
brightnesses fainter than approximately 18 mag. arcsec$^{-2}$ in either 
H or K and 19 in J.

In all three plots, there is an interval where the opacity is constant with 
surface brightness but there is a clear upturn in opacity at the brighter 
values. As the surface brightness limits the accuracy of the SFM 
\citep{Holwerda05d}, these opacity values are also more uncertain. 
However, the result is consistent with studies of inclination effects on 
disks \citep{Giovanelli94,Masters03} and with Freeman's Law \citep{FreemanLaw}.

\begin{figure}[!h]
\includegraphics[width=0.5\textwidth]{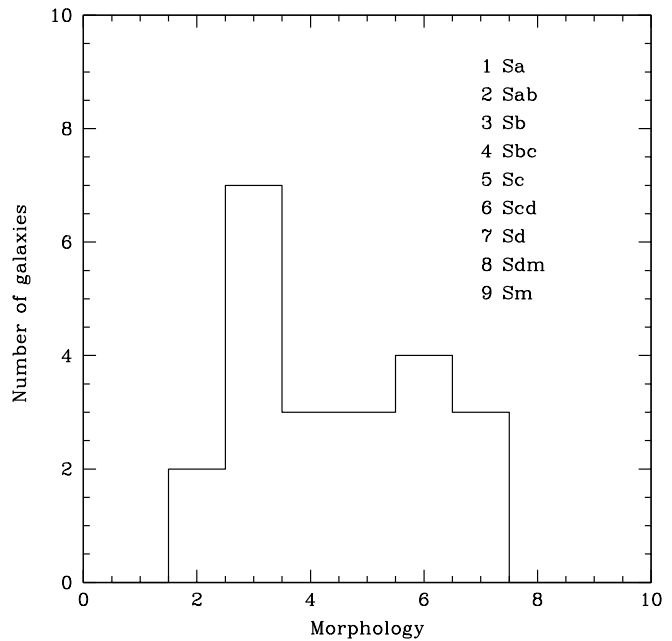}
\caption{\label{MorphHist}The histogram of the Hubble types in this paper's 
sample. The HST Distance Scale Key project selected in favour of late-type 
spirals. However, most spiral galaxy types later then Sab are in our sample.}
\end{figure}

Whether or not the spread in Hubble types (Figure \ref{MorphHist}) has a discernible effect 
on the relation between surface brightness and opacity can be found 
by determining the relations for the early and late type spirals in our sample. 
The relations for early and late types do not appear to be any different in all 
of the 2MASS bands.

\begin{figure}[!h]
\includegraphics[width=0.5\textwidth]{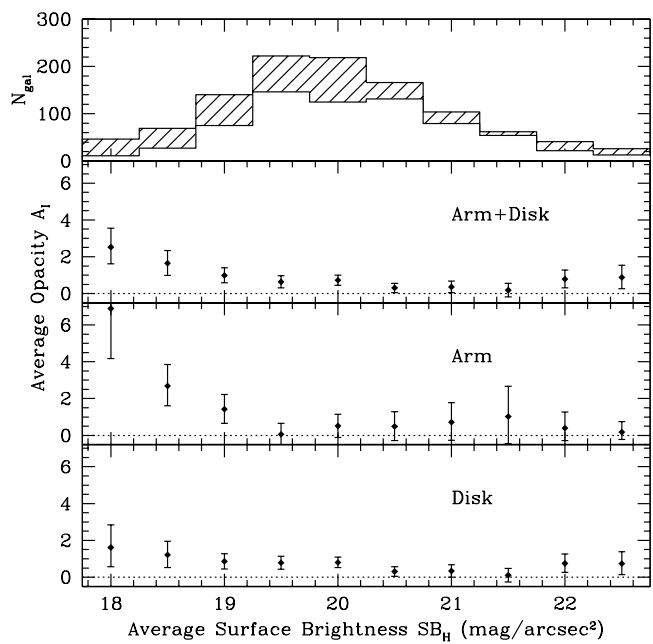}
\caption{\label{SB_h_A_0.5} Top: the number of field galaxies as a function 
of 2MASS surface brightness: the distant galaxies from the science field (solid) 
and the synthetic fields (shaded) without dimming (A = 0). Second from the 
top: the opacity in I (F814W), in magnitudes, as a function of H-band surface 
brightness in 2MASS \citep{2MASS} images. The point at 17.5 mag arcsec$^{-2}$ 
is not based on sufficient statistics for a good comparison. Third from the top: 
the opacity in I as a function of H-band surface brightness, for those regions 
classified as `spiral arm'. Bottom: the opacity in I as a function of H-band 
surface brightness, for those regions classified as disk region, not part of a 
spiral arm.}
\end{figure}

\begin{figure}[!h]
\includegraphics[width=0.5\textwidth]{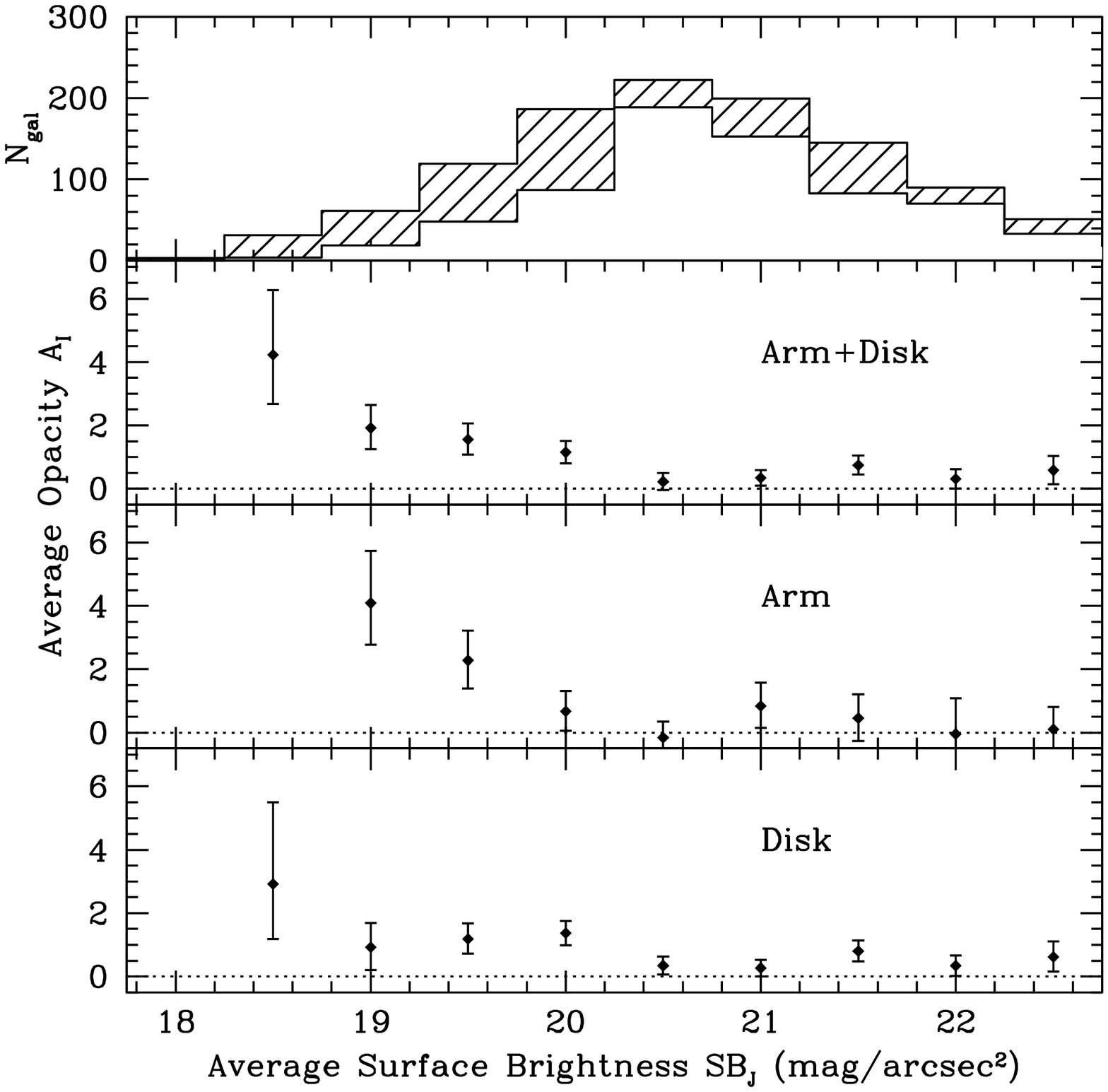}
\caption{\label{SB_j_A_0.5}Top: the number of field galaxies as a function of 
2MASS surface brightness; the distant galaxies from the science field (solid) 
and the synthetic fields (shaded) without dimming (A = 0). Second from the top: 
the opacity in I (F814W), in magnitudes, as a function of J-band surface 
brightness in 2MASS \citep{2MASS} images. Third from the top: the opacity in 
I as a function of J-band surface brightness, for those regions classified as 
`spiral arm'. Bottom: the opacity in I as a function of J-band surface brightness, 
for those regions classified as disk region, not part of a spiral arm.}
\end{figure}

\begin{figure}[!h]
\includegraphics[width=0.5\textwidth]{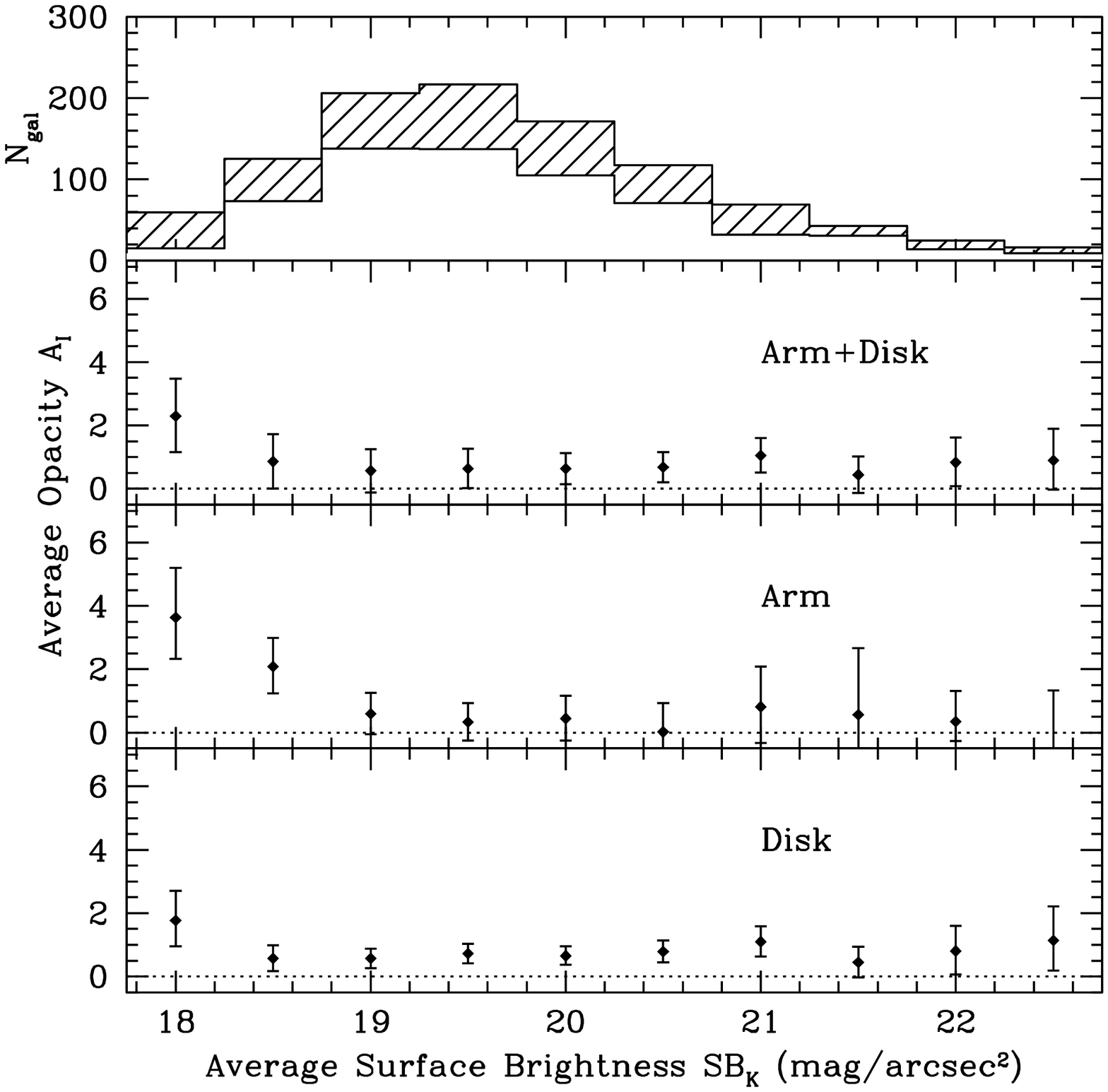}
\caption{\label{SB_k_A_0.5}Top: the number of field galaxies as a function of 
2MASS surface brightness; the distant galaxies from the science field (solid) 
and the synthetic fields (shaded) without dimming (A = 0). Second from the top: 
the opacity in I (F814W), in magnitudes, as a function of K-band surface 
brightness in 2MASS \citep{2MASS} images. Third from the top: the opacity in 
I as a function of K-band surface brightness, for those regions classified as 
`spiral arm'. Bottom: the opacity in I as a function of K-band surface brightness, 
for those regions classified as disk region, not part of a spiral arm.}
\end{figure}

\begin{figure}[!h]
\includegraphics[width=0.5\textwidth]{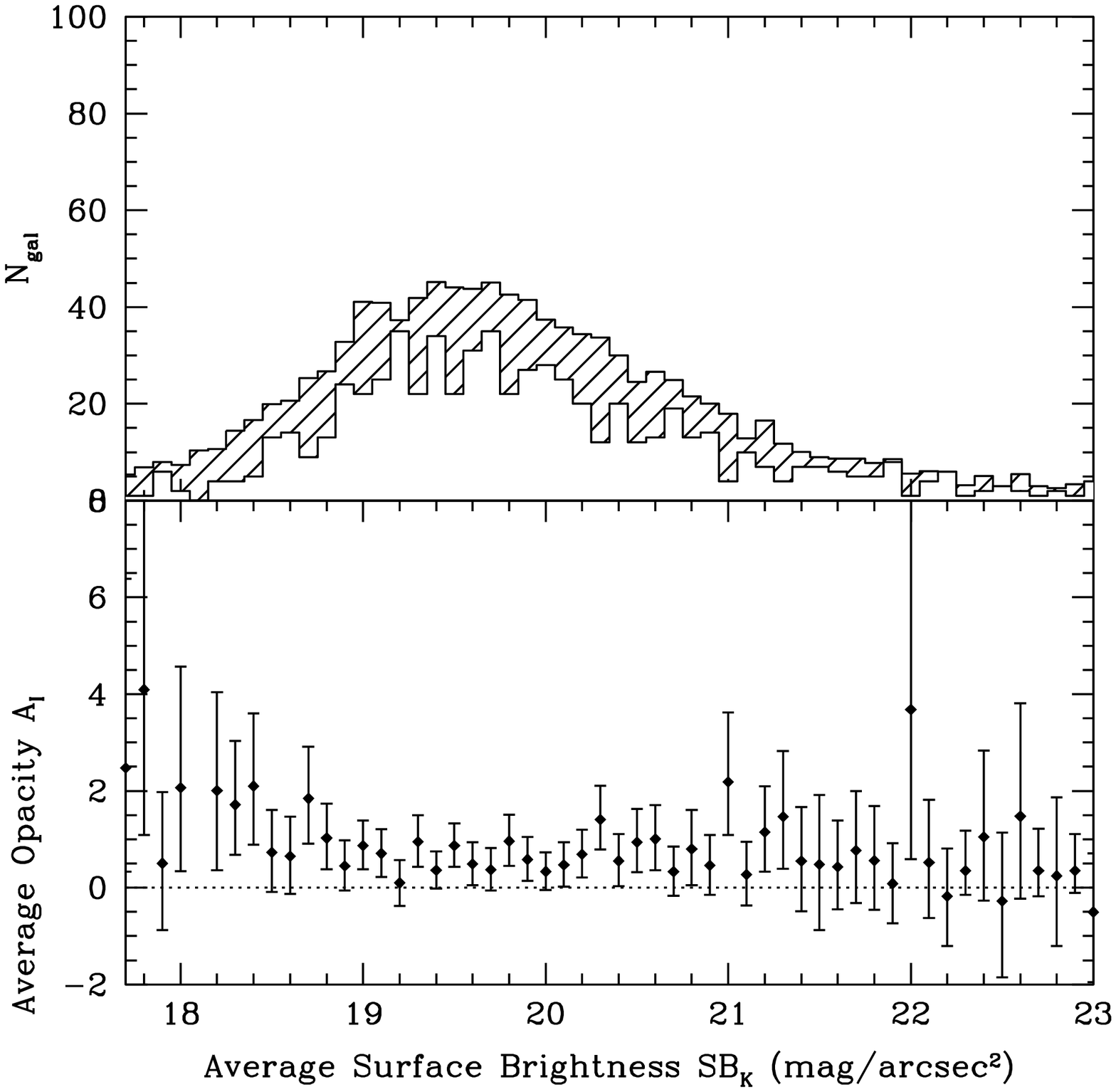}
\caption{\label{SB_k_A_0.1}Top panel: the number of field galaxies as a function 
of 2MASS surface brightness, but at a much smaller sampling scale than Figure 
\ref{SB_k_A_0.5}. The distant galaxies from the science field (solid) and the 
synthetic fields (shaded) without dimming (A = 0). Bottom: the opacity in I (F814W) 
in magnitudes, as a function of K-band surface brightness in 2MASS \citep{2MASS} 
images.}
\end{figure}

\section{\label{secSBAtype}Surface brightness and opacity in arm and disk}

In Figures \ref{SB_h_A_0.5}, \ref{SB_j_A_0.5} and \ref{SB_k_A_0.5}, 
the relation between surface brightness in the H, J, and K bands and 
opacity for the arm and disk regions is also shown. In \cite{Holwerda05b}, 
a relation between the averaged surface brightness and extinction in 
radial annuli was suspected. In the case of spiral arms this relation could 
be steeper than in the rest of the disk. In Figures \ref{SB_h_A_0.5}, 
\ref{SB_j_A_0.5} and \ref{SB_k_A_0.5} the opacities are derived as 
 for arms and disk combined, and separately for the 
sections that were classified as arm or as disk-regions, either inter-arm 
or outside-arm. Classifications of the regions are based on the WFPC2 
mosaic and are described in \cite{Holwerda05a} and \cite{Holwerda05}. 
Opacity measurements can be made for regions fainter than approximately 
18 mag. arcsec$^{-2}$ in either H, J or K, where the brightest disk regions 
are inter-arm regions close to the center. Bright regions are also in 
the middle of spiral arms, notably in star forming regions.

The relation between opacity and surface brightness shown in Figures 
\ref{SB_h_A_0.5}, \ref{SB_j_A_0.5} and \ref{SB_k_A_0.5} seems to be 
dominated by the arm regions in these fields. There is a steep relation 
between opacity and surface brightness in the arms, while there is none 
or perhaps a weak one in the disk regions. Since the measured opacity by the 
SFM is an average for the given area, the higher value for brighter arm 
regions indicates a higher filling factor -or surface density- of molecular 
clouds in these regions. The average values of A are consistent with 
those found by \cite{Valentijn90} but are uncorrected for inclination. 
The brighter regions in the spiral arms are in 
the middle of the spiral arm and near the galaxy's center. This higher 
surface density of dust clouds is consistent with models which 
interpret spiral arms as local overdensities of molecular clouds and 
associated starformation.

\begin{figure}[!h]
\includegraphics[width=0.5\textwidth]{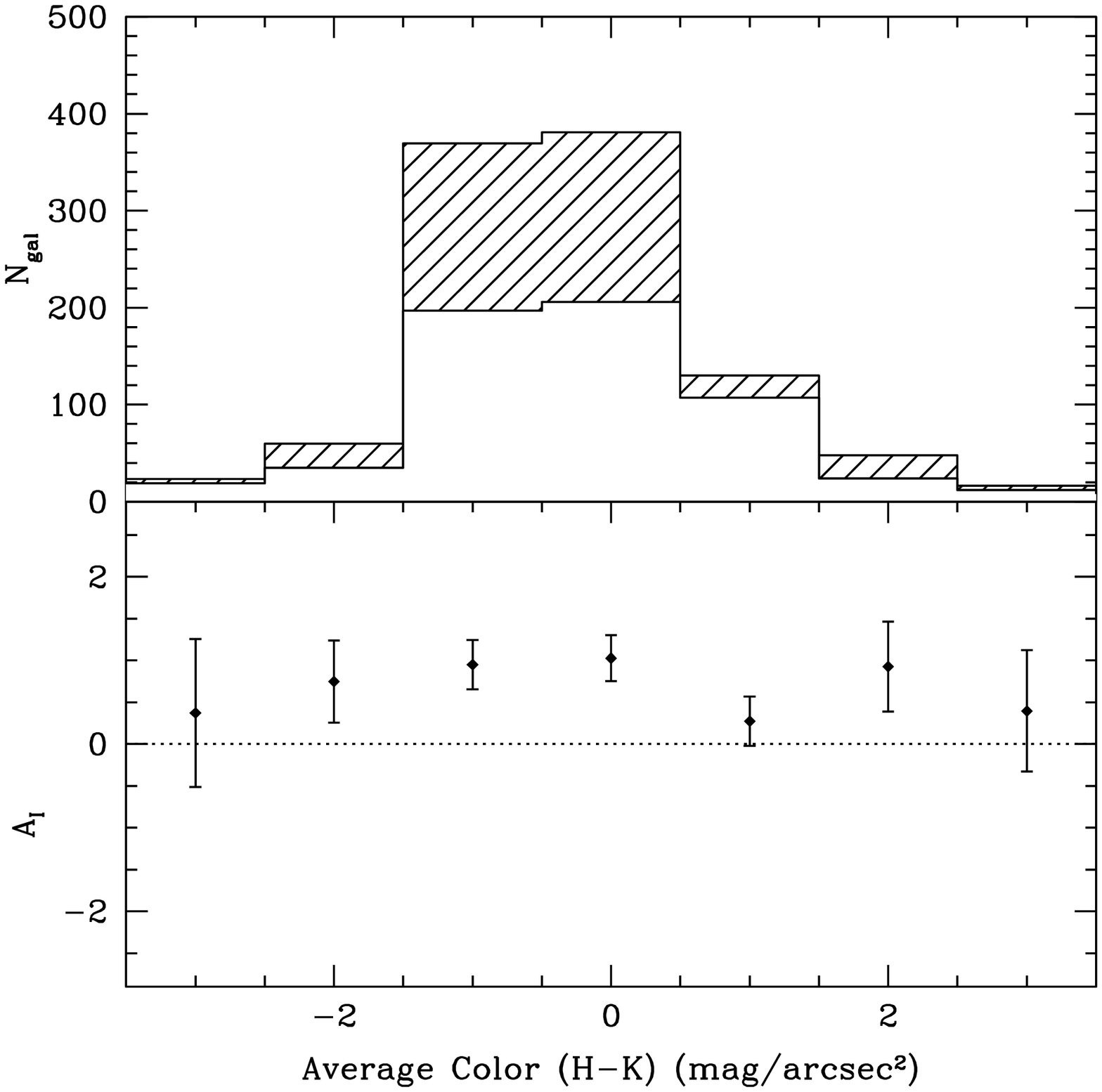}
\caption{\label{Color_H_K_A}The relation between (H-K) color and I-band 
extinction (bottom). Top panel shows the number of distant galaxies for each 
color bin. }
\end{figure}

\begin{figure}[!h]
\includegraphics[width=0.5\textwidth]{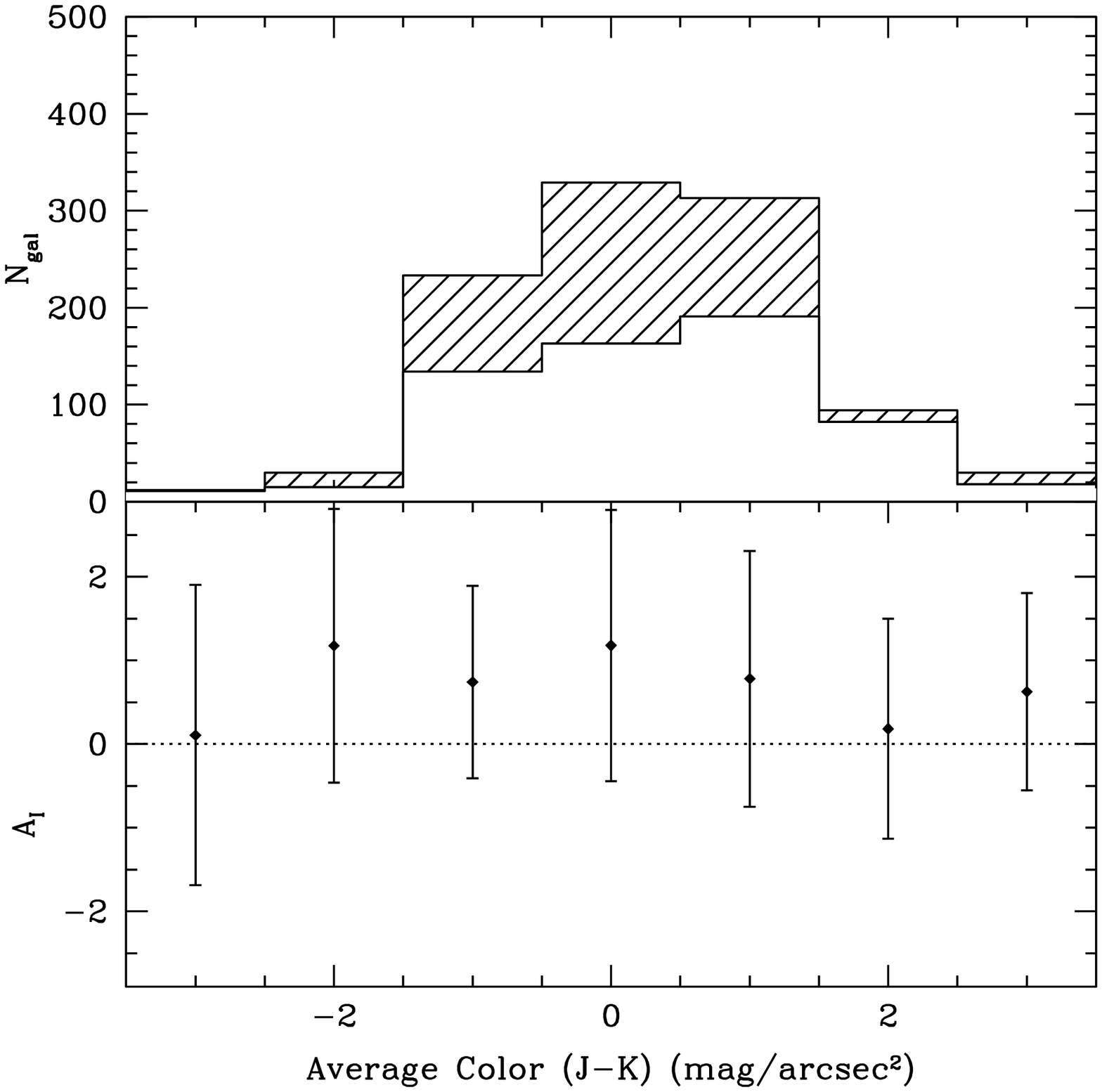}
\caption{\label{Color_J_K_A}The relation between (J-K) color and I-band 
extinction (bottom). Top anel shows the number of distant galaxies for each 
color bin. }
\end{figure}

\section{\label{secColA}Disk opacity and NIR color}

In a similar fashion as the counts of distant galaxies from science and 
synthetic fields are grouped using the disk's surface brightness in 
2MASS images, the color of the foreground disk can be used.
As this color is based on a 2MASS pixel for each distant galaxy, 
the errors are likely to be substantial and it is possible that the distant 
galaxy in the science field itself can influence this color more than 
the surface brightness.

To compensate for the added uncertainty, the sampling of color is 
taken to be 1.0 magnitude. In Figures \ref{Color_H_K_A} and 
\ref{Color_J_K_A}, the opacity in I as a function of the disk's H-K and 
J-K colors are plotted respectively. From these figures we conclude 
that there is very little or no relation between the reddening of the disk 
in the near-infrared bands and the average opacity of the disk in I. 

The motivation for this comparison is the common use of near-infrared 
color as an indicator of stellar mass-to-light ratio. A lack of a strong 
trend of color with opacity would warrant this use. Since the comparison in 
Figures \ref{Color_H_K_A} and \ref{Color_J_K_A} is limited to a 
small range in color, the only conclusion is a lack of a strong relation.

\begin{deluxetable}{l l l l l l}
\tablecaption{\label{table:scalelengths} The scalelengths of the galaxies from the Hubble Distance Scale Key Project. Stellar scalelengths from \cite{Macri05} and dust scalelengths from fitted to the points in \cite{Holwerda05b}.}
\tablewidth{0pt}
\tablehead{
\colhead{Galaxy}	 & \colhead{$R_{typ}$(light)} & \colhead{err}	& \colhead{$R_{typ}$(dust)} & \colhead{err}	& \colhead{$R_{25}$}   \\
				& arcmin.		& & arcmin.	& & arcmin. \\	}

\startdata
   NGC 925   	& 0.588  	& 0.042  	& \nodata 	& \nodata	& 10.47\\
  NGC 1365   	& 0.499  	& 0.012  	& \nodata	& \nodata	& 11.22\\
  NGC 1425   	& 0.292 	& 0.002 	& 10 		& 17  	& 5.75\\
  NGC 2541   	& 0.341  	& 0.023 	& 9 		& 5 	& 6.31\\
  NGC 2841   	& 0.358 	& 0.006   	& -12  	& \nodata	& 8.13\\
  NGC 3198   	& 0.290  	& 0.015 	& 3.5  	& \nodata	& 8.51\\
  NGC 3319   	& 0.323	& 0.046 	& 6.5 	& 2.0  	& 6.17\\
  NGC 3351   	& 0.391  	& 0.033   	& -14.1 	& 4.0  	& 7.41\\
  NGC 3621   	& 0.410  	& 0.029  	& \nodata	& \nodata	& 12.3\\
  NGC 3627   	& 0.450  	& 0.011  	& \nodata	& \nodata	& 9.12\\
  NGC 4414   	& 0.258  	& 0.032 	& 6 		& 17  	& 3.63\\
  NGC 4535   	& 0.360  	& 0.014 	& 8.4	 	& 2.0  	& 6.17\\
  NGC 4536   	& 0.305 	& 0.007  	& \nodata	& \nodata	& 7.59\\
  NGC 4548   	& 0.240 	& 0.010 	& 9.1 	& 4.8  	& 5.37\\
  NGC 4639   	& 0.150 	& 0.009 	& 1.1   	& 0.4  	& 2.75\\
  NGC 4725   	& 0.562 	& 0.008 	& 9.8  	& \nodata	& 10.72\\
  NGC 7331   	& 0.509  	& 0.006 	& 15  	& \nodata	& 10.47\\
\enddata
\end{deluxetable}

\begin{figure}[!h]
\includegraphics[width=0.5\textwidth]{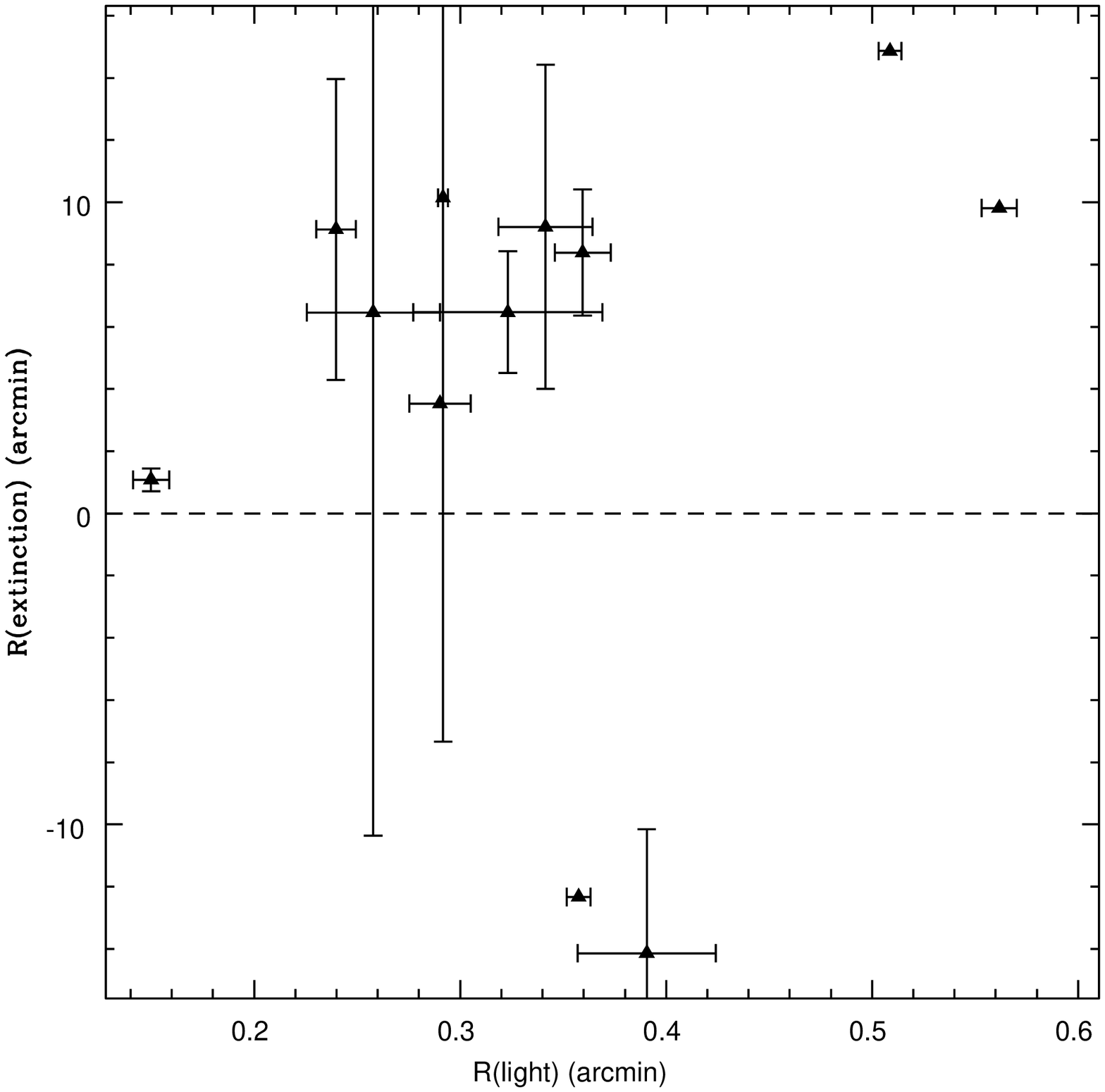}
\caption{\label{scalelengths}The scalelengths of extinction and light in the I 
band for those galaxies in both our sample that of \cite{Macri00}. Most of 
scalelengths of the dust are much larger than those of the light, contrary to 
earlier results but a few are negative.  }
\end{figure}

\section{\label{secScales}Dust and light scalelengths}

A subject of interest is the extent of dust in spiral disks. In \cite{Holwerda05b}, 
we presented radial profiles of individual WFPC2 fields as well as 
averages over Hubble type and arm/disk regions. A common indicator 
of disk scale is the exponential scalelength of the radial light profile. 
\cite{Macri00,Macri05} present photometry and exponential disk fits 
on the Distance Scale Key Project spiral galaxies in order to provide 
a calibration for the Tully-Fisher relation. They present photometric 
diameters, scalelengths and central surface brightnesses in B,V,R 
and I for a large subset of our sample. 

A simple exponential disk was fitted to the radial extinction measurements 
of individual fields presented in Table 3 of \cite{Holwerda05b}. These 
values are poorly determined as the small field-of-view of the WFPC2 
results only in a few opacity measurements per galaxy. Some of the 
extinction profiles rise, rather than decline. This is the result of the 
presence of spiral arms in the relevant WFPC2 images. 

Figure \ref{scalelengths} compares the scalelengths for the light and 
extinction in I for individual galaxies. The scalelengths and the $R_{25}$ 
\citep{RC3} are also listed in Table \ref{table:scalelengths}. 
These scale-lenghts of the opacity are an order of magnitude larger 
than the scale-lengths of the stellar light. This is a direct result of the 
very gradual decline with radius seen in \cite{Holwerda05b}.

This would be in contradiction to results from edge-on galaxies on dust 
scalelengths which put it at 1.4 times the scale-length of the stars 
\citep{Xilouris99,Radovich01} but not inconsistent with sub-mm 
observations which put the scale of the dust disk somewhere between 
the HI and stellar scales \citep{Alton98, Davies99, Trewhella00}.

One explanation for our result could be that the Cepheid distance 
project pointed the HST at the arms of these galaxies, biasing the 
extinction profile. However, it seems likely that a dark, cold cloud 
component displays a different relation with radius than that obtained 
from previous measurements of dust via emission (warm dust) or 
stellar reddening (diffuse dust). This is consistent with the picture 
of the ISM of spiral disks emerging since the first results by \cite{Valentijn90} 
(e.g. \cite{Block94b} and FIR observations \cite{Nelson98,Alton98b,Trewhella00,Popescu02,Hippelein03}) 
In addition, the general assumption that the dust is distributed as an 
exponential disk might need to be revisited. 

Future work with FIR/sub-mm observations or counts of 
background galaxies should characterise the scale and radial 
profile of dust disks in spiral galaxies. An improved measurement 
of the dust scalelength from galaxy counts would require a larger 
solid angle per individual galaxy.

\begin{deluxetable}{l l l l l}
\tablewidth{0pt}
\tablecaption{\label{table_keyproj}Opacity of the WFPC2 field from Cepheid reddening and Galaxy counts.}
\tablehead{
\colhead{Galaxy}	 & \colhead{E(V-I)} & \colhead{$\sigma$ E(V-I)} & \colhead{$A_{Cepheid}$} & \colhead{$A_{SFM}$ }\\}

\startdata
NGC925	 	& 0.21	 & 0.02	 & 0.8	 & $-0.4_{-0.3}^{+0.3}$\\
NGC1365	 	& 0.20	 & 0.02	 & 0.8	 & $0.5_{-0.3}^{+0.3}$\\
NGC1425	 	& 0.16	 & 0.03	 & 0.6	 & $0.5_{-0.3}^{+0.3}$\\
NGC2541	 	& 0.20	 & 0.02	 & 0.8	 & $0.8_{-0.3}^{+0.3}$\\
NGC3198	 	& 0.15	 & 0.04	 & 0.6	 & $0.8_{-0.3}^{+0.3}$\\
NGC3319	 	& 0.13	 & 0.04	 & 0.5	 & $0.9_{-0.4}^{+0.4}$\\
NGC3351	 	& 0.24	 & 0.04	 & 1.0	 & $1.2_{-0.6}^{+0.5}$\\
NGC3621-OFF	& 0.36	 & 0.04	 & 1.4	 & $1.0_{-0.4}^{+0.3}$\\
NGC3627	 	& 0.24	 & 0.03	 & 1.0	 & $2.1_{-0.7}^{+0.7}$\\
NGC4321	 	& 0.22	 & 0.03	 & 0.9	 & $2.3_{-0.8}^{+0.7}$\\
NGC4414-2	& 0.15	 & 0.04	 & 0.6	 & $0.7_{-0.4}^{+0.3}$\\
NGC4496A	& 0.14	 & 0.01	 & 0.6	 & $5.0_{-0.9}^{+0.8}$\\
NGC4535	 	& 0.19	 & 0.02	 & 0.8	 & $0.7_{-0.4}^{+0.4}$\\
NGC4536	 	& 0.18	 & 0.02	 & 0.7	 & $0.9_{-0.4}^{+0.4}$\\
NGC4548	 	& 0.18	 & 0.04	 & 0.7	 & $0.8_{-0.4}^{+0.3}$\\
NGC4639	 	& 0.12	 & 0.04	 & 0.5	 & $0.8_{-0.3}^{+0.3}$\\
NGC4725	 	& 0.29	 & 0.03	 & 1.2	 & $0.8_{-0.3}^{+0.3}$\\
NGC7331	 	& 0.25	 & 0.05	 & 1.0	 & $0.3_{-0.3}^{+0.3}$\\
\enddata

\end{deluxetable}

\begin{figure}[!h]
\includegraphics[width=0.5\textwidth]{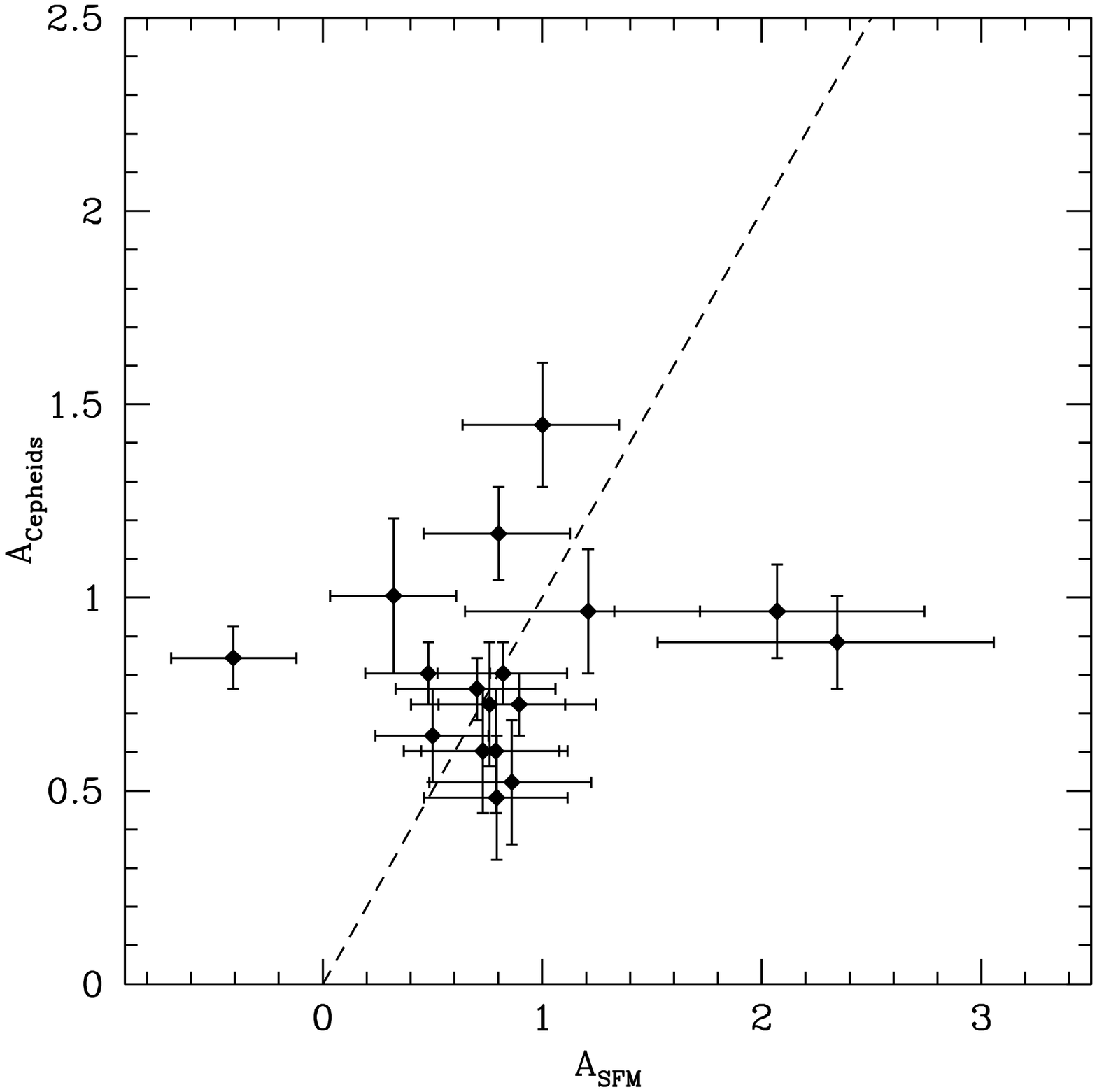}
\caption{\label{KeyProj}The average extinction derived from the number of 
distant galaxies compared to the extinction derived from the Cepheid 
reddening (E(V-I)) from \cite{Keyproject}. There is no clear linear relation 
(dashed line). The Cepheid extinctions saturate at higher opacities as the 
Distance Scale project selected against high-extinction Cepheids. Note 
that cosmic variance in the number of background galaxies can produce 
a negative opacity value.}
\end{figure}

\section{\label{secKeyProj}Comparison to Cepheid Reddening}

\cite{Keyproject} present reddening values (V-I) based on the 
photometry of the Cepheids in the Distance Scale Key Project 
galaxies. The reddening E(V-I) was converted to an opacity in I 
using the Galactic Extinction Law.  In Table \ref{table_keyproj} 
and Figure \ref{KeyProj} the average extinction values from number 
counts for the combined Wide Field chips in each galaxy and the 
average extinction derived from the Cepheid reddening are 
compared. The comparison is not a straightforward one. The 
extinction in front of a Cepheid variable is local and biased 
against high extinction values. \cite{Keyproject}, however, gave 
average reddening based on all Cepheids in the field. The 
extinction from the number of distant galaxies is for the entire 
height of the disk and an average for all the chips. 

In Figure \ref{KeyProj}, the scatter in $A_{Cepheid}$ for a given 
$A_{SFM}$ value can be explained by variations in the average 
depths of the Cepheids in the disks; the lack of high $A_{Cepheid}$ 
values is probably the result of the detection bias for Cepheids. 
The opacity from the number of distant galaxies can probe much 
higher average extinction, something the Cepheid reddening 
selects against.

\section{\label{secML}Discussion: implication for our view of spiral disk}

The primary result presented in this paper is the relation between 
near-infrared surface brightness and disk opacity. 
Such a relation is consistent with the observation that brighter galaxies 
are also more opaque (e.g. \cite{Masters03}). It is also consistent with 
``Freeman's Law'' \citep{FreemanLaw}): the central surface 
brightness of the disk is constant, regardless of the inclination.
A direct relation between surface brightness and disk extinction has some 
ramifications for photometric measurements of spiral disks. 
The slope of the light profile of a spiral disk is underestimated as more light 
is hidden by dust in the brighter parts. As a result, the scalelength of the 
exponential disk is overestimated. This has implications for 
dynamical models of spiral disks. If the stellar disk is in fact somewhat 
more compact than observed, the stellar mass contribution of this disk 
is greater in the center. 
To model rotation curves, it is common to assume a ``maximum disk'': 
the light profile is converted to a mass distribution using the maximum 
conversion factor allowed by the rotation curve.
If surface brightness and extinction are related, as Figures \ref{SB_h_A_0.5}, 
\ref{SB_j_A_0.5} and \ref{SB_k_A_0.5} indicate, then the stellar mass estimate 
is underestimated in the center of the disk. The stellar profile, corrected for 
extinction, would mitigate the need for dark matter in the center of spiral disks. 
A stellar mass profile, corrected for extinction, would reach the point of 
``maximum disk'' at a slightly shorter radius and at a lower M/L conversion 
factor (see also \cite{Gonzalez91}). A different M/L ratio for the ``maximum disk'' 
would imply a slope of the Tully-Fisher relation, slightly lower than 3.5, 
according to \cite{BelldeJong}. 

The second result in this paper is the lack of a strong relation between 
surface brightness and opacity in the  part of the disk outside the spiral 
arms. This constant opacity value for the disk is consistent with the flat 
average radial profile for the disk regions, that we found from the same 
data \citep{Holwerda05b}. However, this result is predominantly limited 
to the optical disk of the galaxies. The constant opacity as a function of 
surface brightness and radius suggests to us that this component 
may extend to a point beyond the optical radius of the disk. 
Evidence for a cold disk with warmer dust in the arms has been found 
from emission \citep{Popescu02, Hippelein03}.
However, the actual extent of this dust component can be explored with 
counts of galaxies in many fields of a single disk -instead of an average over several disks. 

The spiral arms show a stronger relation between surface brightness and 
opacity. This would be consistent with the view that the arms are overdensities in the disk. 
The fact that the relation between surface brightness and opacity is different 
for arm and disk regions, implies different dust components. As a result, a single 
exponential disk may be an oversimplification of the general dust distribution.

The third result is that no strong relation between NIR color and dust opacity 
seems to be present. This is consistent with the results of \cite{BelldeJong}. 
They find that a good indicator of the M/L in a spiral disk is its near-infrared 
color. They also discuss effects of dust on their models but the effect of dust 
reddening and dimming is estimated to cancel out to first order. 
\cite{BelldeJong} advise against using their M/L values for 
anything but a whole disk, as local extinction is likely to be 
patchy and more grey in nature. In fact, \cite{BelldeJong} already allow for 
sub-maximum disks as a shift in their zero-point in the relation 
between M/L and color. Recent work on rotation curves of spiral 
disks, such as \cite{Kassin05}, should therefore not be affected by 
the use of the color for the M/L indicator, provided the profile is determined 
predominantly outside the spiral arms.

\section{\label{secSBconcl}Conclusions}

The main conclusions we can draw from the numbers of distant 
galaxies seen through a spiral disk, as analysed in this paper are:
\begin{itemize}
\item[1.] The dust opacity of the spiral disks increases with increasing surface 
brightness of the disk (Figures \ref{SB_h_A_0.5}, \ref{SB_j_A_0.5} and \ref{SB_k_A_0.5}).
\item[2.] This effect is mainly restricted to the spiral arms. This implies 
a higher surface density of clouds in the spiral arms. (Figures 
\ref{SB_h_A_0.5}, \ref{SB_j_A_0.5} and \ref{SB_k_A_0.5}).
\item[3.] For the disk regions, the opacity is constant with surface 
brightness (Figures \ref{SB_h_A_0.5}, \ref{SB_j_A_0.5} and \ref{SB_k_A_0.5}).
\item[4.] The dust opacity does not strongly correlate with the near-infrared 
color of the foreground disk (Figures \ref{Color_H_K_A} and \ref{Color_J_K_A}).
\item[5.] An exponential disk appears to be a poor description of the dust distribution 
as evident from the much larger values compared to the stellar scalelength 
(Figure \ref{scalelengths}) but better counts on single galaxies should should 
provide a better constraint.
\item[6.] The average values of the Cepheid reddening and the disk 
opacity correspond reasonably well for the lower extinction values. 
High disk extinction does not show in the Cepheid reddening, probably 
due to selection effects (Figure \ref{KeyProj}).
\end{itemize}

If surface brightness and disk opacity are linked, fitting a maximum M/L 
to the light distribution to model the stellar dynamical component 
(``Maximum Disk") is an unrealistic approach. In addition, this could 
explain the results from \cite{Giovanelli94} and \cite{Masters03} and 
``Freeman's Law'' as well. The grey behaviour of the opacity as measured 
from numbers of distant background galaxies can be explained by the 
fact that the observable distant galaxies are on low extinction lines-of-sight. 
The fact that the opacity is related to the surface brightness of the disk 
does pose a problem for the use of a single M/L value in dynamical fits.

The Advanced Camera for Surveys on Hubble has imaged many more 
galaxies and the presented relation between surface brightness and 
extinction could be further explored using these more recent data. 
Any systematic effect of combining counts from several different 
disks can be avoided altogether in a similar analysis performed on a single disk.
However, as \cite{Gonzalez03} and \cite{Holwerda05d} point out, the SFM is limited 
to the less crowded parts of the disk and the relation between extinction 
and surface brightness could only be extended at the fainter end. 

\acknowledgements
The authors would like to thank the referee, Edwin Valentijn, for his 
insightful comments on the paper and very useful suggestions and 
Roelof de Jong and Susan Kassin for discussions on these results. 
The authors would especially like to thank Lucas Macri for making 
his results available for comparison, prior to their publication.

This research has made use of the NASA/IPAC Extragalactic Database, 
which is operated by the Jet Propulsion Laboratory, California Institute of 
Technology, under contract with the National Aeronautics and Space 
Administration (NASA). This work is based on observations with the 
NASA/ESA Hubble Space Telescope, obtained at the STScI, which is 
operated by the Association of Universities for Research in Astronomy 
(AURA), Inc., under NASA contract NAS5-26555. In addition, this 
publication makes use of the data products from the Two Micron All Sky 
Survey, which is a joint project of the University of Massachusetts and 
the Infrared Processing and Analysis Center/California Institute of 
Technology, funded by the National Aeronautics and Space 
Administration and the National Science Foundation.
Support for this work was provided by NASA through grant number 
HST-AR-08360 from STScI. STScI is operated by AURA, Inc., under 
NASA contract NAS5-26555. We are also grateful for the financial 
support of the STScI DirectorÕs Discretionary Fund (grants 82206 and
82304 to R. J. Allen) and of the Kapteyn Astronomical Institute of 
Groningen University.

\bibliographystyle{astron} 
\bibliography{./SBpaper}

\end{document}